%%%%%%%%%%%%%%%%%%%%%%%%%%%%%July 9, 1999%%%%%%%%%%%%%%%%%%%%%%%%%%%%%%%%%%%%%%%%%%%%
\magnification = \magstep 1
\overfullrule=0pt
\font\twelvebf=cmbx12
\font\ninerm=cmr9
\nopagenumbers
\overfullrule=0pt
\line{\hfil RU 99-9-B}
\line{\hfil KIAS-P99054}
\line{\hfil CCNY-HEP 99/3}
%\line{\hfil }
\vskip .3in
\centerline{\twelvebf Gauge invariant variables and the Yang-Mills-Chern-Simons theory}
\centerline{\twelvebf}
\vskip .3in
\baselineskip=14pt
\centerline{\ninerm DIMITRA KARABALI}
\vskip .05in
\centerline{\ninerm Department of Physics and Astronomy}
\centerline{\ninerm Lehman College of the City University of New York}
\centerline{\ninerm New York, NY 10468}
%\vskip .1in
\centerline{\ninerm and}
%\vskip .1in
\centerline{\ninerm Physics Department, Rockefeller University}
\centerline{\ninerm New York, New York 10021}
\centerline{\ninerm karabali@theory.rockefeller.edu}
\vskip .3in
\centerline{\ninerm CHANJU KIM}
\vskip .05in
\centerline{\ninerm Korea Institute for Advanced Study}
\centerline{\ninerm 130-012 Seoul, Korea}
\centerline{\ninerm cjkim@kias.re.kr}
\vskip .3in
\centerline{\ninerm V.P. NAIR}
\vskip .05in
\centerline{\ninerm Physics Department}
\centerline{\ninerm City College of the City University of New York}
\centerline{\ninerm New York, New York 10031}
\centerline{\ninerm vpn@ajanta.sci.ccny.cuny.edu}
\vskip .3in
\baselineskip=14pt
\centerline{\bf Abstract}

A Hamiltonian analysis of Yang-Mills (YM) theory in (2+1) dimensions with a level $k$
Chern-Simons term is carried out using
a gauge invariant matrix parametrization of the potentials. 
The gauge
boson states are constructed and the contribution of the dynamical
mass gap to the gauge boson mass is obtained.
Long distance properties of vacuum expectation values are 
related to a Euclidean
two-dimensional YM theory coupled to $k$ flavors of Dirac fermions in the
fundamental representation. We also discuss the
expectation value of the Wilson loop operator and give a
comparison with previous results.
\vfill\eject

\footline={\hss\tenrm\folio\hss}
\def\bp {\bar p}

\def\bG {\bar{G}}
\def\bA {\bar{A}}
\def\bB {\bar{B}}
\def\bE {\bar{E}}
\def\bPi {\bar{\Pi}}
\def\bx {\bar{x}}
\def\by {\bar{y}}

\def\vx {{\vec x}}

\def\vy{\vec{y}}
\def\vv {\vec{v}}
\def\vu {\vec{u}}
\def\vw {\vec{w}}

\def\dag {\dagger}
\def\del {\partial}
\def\bdel{\bar{\partial}}

\def\e {\epsilon}
\def\d {\delta}

\def\half {{\textstyle {1 \over 2}}}
\def\bG {\bar{G}}

\def\H {{\cal H}}

\def\G {{\cal G}}

\def\S {{\cal S}}
\def\ra {\rangle}
\def\la {\langle}
\def\Tr {{\rm Tr}}

\def\bD {{\bar D}}
\def\bQ {{\bar Q}}
\def\bA {{\bar A}}
\def\bJ {{\bar J}}

\baselineskip =16pt
\noindent{\bf 1. Introduction}
\vskip .1in 
In recent papers, we have carried out a nonperturbative analysis of
Yang-Mills (YM) gauge theories in two spatial dimensions in a Hamiltonian
formulation [1-3].
The use of a matrix parametrization for the gauge potentials enabled us to
derive a formulation in terms of gauge invariant variables. This framework and the 
use of some techniques
from
conformal field theory helped to simplify the analysis of the spectrum of the theory, 
leading to results on the
mass
gap and the vacuum wavefunction. A recent analytic calculation of 
the string
tension
gave values within 3\% of the values from Monte Carlo simulations.

In this paper, we extend our gauge invariant Hamiltonian analysis to YM theory 
with a Chern-Simons (CS) mass
term
added, i.e., the YMCS theory. In particular we shall focus on how the pure YM 
results on the mass gap, the vacuum wavefunction and the expectation value of Wilson loop
are affected by the addition of the CS term. 

It is well known, that the presence of the CS term generates a perturbative mass 
for the gauge bosons [4]. The question of interest here is whether and how
the perturbative mass gets augmented or modified by the dynamical generation of mass,
which is
known to occur for the pure YM case.
Our approach to this problem is the construction of gauge invariant states 
corresponding to the
dynamical gauge bosons and the analysis of the corresponding Schr\"odinger equation. 
The situation is essentially
analogous to our analysis in the case of pure YM case.

There have, of course, been many papers
analyzing
different aspects of the YMCS theory. 
A particular way of constructing the lower excited states of the YMCS 
theory has been given by Grignani {\it et al} [5].
Later we shall comment on the extent to which our results differ from this work.

In section 2, the reduction to gauge invariant variables is carried out. Some of
the eigenstates of the kinetic energy operator are obtained in section 3.
Section 4 deals with the inclusion of the potential energy, vacuum wavefunction, 
screening, etc.
In section 5, a comparison to other recent work is made. The paper concludes with a short
discussion.
\vskip .1in
\noindent{\bf 2. Gauge invariant variables for YMCS}
\vskip .1in
We shall discuss an $SU(N)$-gauge theory. The gauge potentials can be written as
$A_\mu = -i t^a A_\mu ^a$, $\mu =0,1,2$, where $t^a$ are hermitian 
$(N \times N)$-matrices which form a basis of the Lie algebra of $SU(N)$ with
$[t^a, t^b ] = i f^{abc} t^c,~~{\rm {Tr}} (t^at^b) = {1 \over 2} \delta ^{ab}$.
The action for the YMCS theory is given by $S =S_{YM}+S_{CS}$, with
$$\eqalign{
S_{YM}&= {1\over 4e^2}\int d^3x~ F_{\mu\nu}^a F^{a\mu\nu}\cr
S_{CS}&= -{k\over 4\pi} \int d^3x~\Tr \left( A_\mu\partial_\nu A_\alpha 
+{2\over 3} A_\mu A_\nu A_\alpha \right)\e^{\mu\nu\alpha}\cr
F_{\mu\nu}^a&= \partial_\mu A_\nu^a - \partial_\nu A_\mu^a +f^{abc}A_\mu^b
A_\nu^c\cr}
\eqno(1)
$$
where $e$ is the coupling constant; $e^2$ has the dimension of mass. 
The parameter $k$ is an integer, the level number
of the CS term. {}From now on we
shall work in the
$A_0 =0$ gauge, which is convenient for a Hamiltonian formulation.

The canonical momenta are easily identified and the electric field operators
are given by
$$\eqalign{
E^a&= {\Pi^a\over 2}+{ik\over 8\pi} A^a= -{i\over 2}{\delta \over \delta \bA^a}
+{ik\over 8\pi}A^a\cr
\bE^a&= {{\bar \Pi}^a\over 2}-{ik\over 8\pi} \bA^a= -{i\over 2}{\delta \over
\delta A^a}
-{ik\over 8\pi}\bA^a\cr}\eqno(2)
$$
where we use complex components, $E= \half (E_1+iE_2),~A=\half (A_1+iA_2)$,
$E_i= { 1 \over e^2} F_{0i}$. The commutation rule for
$E^a,\bE^a$ is given by
$$ 
[\bE^a(\vx) , E^b(\vy)]= {k\over 8\pi} \delta^{ab}\delta (\vx-\vy)\eqno(3)
$$ 

The Hamiltonian can be written as
$$\eqalign{
\H&= T+V\cr
T=2e^2\int d^2x ~E^a\bE^a,~&~~~~~~V={1\over 2e^2} \int d^2x~B^aB^a\cr
B^a&= {1\over 2} \epsilon_{ij} F^a_{ij}\cr}\eqno(4)
$$
We have normally ordered the kinetic energy operator $T$ in accordance with (3).

The Gauss law operator $I^a$ is given by
$$
I^a= (D{\bar \Pi}+\bD \Pi )^a +{ik\over 4\pi} (\partial \bA^a -\bdel
A^a)\eqno(5)
$$
The physical states must obey the condition $\int \theta^a (\vx)I^a(\vx) \vert \psi
\ra =0$, for
functions $\theta^a (\vx)$ which vanish at spatial infinity. This
requirement, along with equations (2-5), will define the theory.

In carrying out a similar analysis for the pure Yang-Mills case
[1-3]
we used a particular matrix parametrization of the gauge fields, which
eventually led
to a gauge invariant formulation. We shall use the same parametrization here,
namely
$$
A= -\partial M~M^{-1}~~, ~~~~~~~~~~~~~\bA = M^{\dagger -1}\bdel M^\dagger
\eqno(6)
$$
where $M$ is
a complex
$SL(N,{\bf C})$-matrix. The wavefunctions in the $A$-diagonal
representation
are thus functions of $M,M^\dagger$. {}From (6) it is clear that the gauge transformation
$A_i\rightarrow A_i^h= h A_i h^{-1} -\partial_i h ~h^{-1}$, is expressed in
terms of
$M,~M^\dagger$ by $ M\rightarrow hM,~M^\dagger \rightarrow M^\dagger h^{-1}$ for
$h(x)\in SU(N)$. Since the Gauss law operator in (5) generates gauge
transformations, we
see that the Gauss law condition is equivalent to
$$
\Psi (hM,M^\dagger h^{-1})= \left[ 1+ {k\over 2\pi} \int \Tr \left( M^{\dagger
-1}\bdel
M^\dagger \partial \theta +\bdel \theta \partial M M^{-1}\right)\right]~\Psi
(M,M^\dagger
)\eqno(7)
$$
where $h(x)\approx 1+\theta (x)$, $\theta =-it^a\theta^a$.
The general form of the wavefunction obeying (7) can be written as
$$ \eqalign{
\Psi (M,M^\dagger ) & = \exp\left[ {k\over 2}\left( \S (M^\dagger )- \S
(M)\right)\right]~
\chi (H) \cr
& \equiv e^ {i \omega (M, M^{\dag})} \chi (H) \cr}
\eqno(8)
$$
where $\chi$ is gauge invariant and depends on $M,M^\dagger$ only via the
gauge invariant
combination $H=M^\dagger M$. $\S (M)$ is the Wess-Zumino-Witten (WZW) action for
$M$ given by [6]
$$
\S (M) ={1\over 2\pi} \int d^2x~ \Tr (\partial M \bdel M^{-1}) +{i\over 12\pi}
\int \Tr (M^{-1}dM)^3 \eqno(9)
$$
By virtue of the Polyakov-Wiegmann (PW) identity [7]
$$
\S (hM) = \S (h) +\S (M) -{1\over \pi} \int d^2x~ \Tr (h^{-1}\bdel h \partial
M~M^{-1})
\eqno(10)
$$
it is easily checked that (8) is the general solution to the Gauss law
condition (7).

We have previously calculated the volume measure for gauge invariant
configurations [1,8]
as
$$
d\mu ({\cal C}) = d\mu (H) e^{2c_A \S (H) }\eqno(11)
$$
where ${\cal{C}}$ is the space of gauge potentials modulo gauge transformations,
$d\mu (H)$ is the product of the Haar measure for $H$ over all space
points
and $c_A$ is the quadratic Casimir for the adjoint representation of $SU(N)$.
In calculating the inner product of two states, notice that since the phase 
$\omega$ in (8) is real, $\Psi_1^* \Psi_2 = \chi_1^* \chi_2$ is gauge invariant.
Hence the inner  
product for two states with wavefunctions of the form (8)
is given by
$$
\la 1\vert 2\ra = \int d\mu (H) e^{2c_A \S (H)} \chi_1^* \chi_2
\eqno(12)
$$

In computing matrix elements of operators involving $E,~\bE$, the phase $e^{i
\omega}$ does contribute since $E,~\bE$ do not commute with it. We can however use
$\chi (H)$ as the wavefunction of the state (with the inner product (12))
provided every operator ${\cal{O}}$ acting on $\Psi$ is redefined as $e^{-i
\omega} {\cal{O}} e^{i \omega}$ in terms of its action on $\chi$ 's.

In terms of the $\chi$ 's, the corresponding Schr\"odinger equation reads
$$
{\cal{H}}' \chi (H) = E \chi (H)
\eqno(13)
$$
where ${\cal{H}}' = e^{-i \omega} {\cal{H}} e^{i \omega} = e^{-i \omega} T
e^{i \omega} + V = T' + V $. ${\cal{H}}'$ can be expressed in terms of the
gauge invariant variable $H=M^{\dag} M$. 
$$\eqalign{
T' & ={e^2 \over 2} \int H_{ab} (\vx) \left( \int_y \bar{\cal{G}}(\vx,\vy) \bp (\vy) -
{ik
\over 4\pi} (\del H H^{-1})(\vx) \right)_a \left( \int_u {\cal{G}}(\vx,\vu) p(\vu) + {ik
\over
4\pi} (H^{-1}
\bdel H) (\vx)\right)_b \cr
V & = {2 \over e^2} \int \bdel (\del H H^{-1})_a \bdel (\del H H^{-1})_a \cr }
\eqno(14)
$$
In arriving at (14) we have used the fact that with the parametrization (6) of
the potentials, we have
$$ \eqalign{
\Pi ^a (\vx) &= -i {\d \over {\d\bA ^a (\vx)}} = i M^\dagger _{ba} (\vx) \int _y 
\bG (\vx,\vy) \bp
_b 
(\vy) \cr
\bPi ^a (\vx)&= -i {\d \over {\d A ^a (\vx)}} = -i M _{ab} (\vx) \int _y G (\vx,\vy) p _b
(\vy) \cr} \eqno(15)
$$
where $M_{ab} = 2 Tr (t^a M t^b M^{-1})$ is the adjoint representation of $M$.
(Similarly $H_{ab} = 2 Tr (t^a H t^b H^{-1})$ is the adjoint representation of
$H$.)
$G (\vx,\vy),~ \bG(\vx,\vy)$ are the Green's functions for the operators $\del~,~\bdel$
respectively.
$$\eqalign{
& \del _x G(\vx,\vy) = \bdel_x \bG (\vx,\vy) = \d (\vx-\vy) \cr
& G(\vx,\vy) = { 1 \over {\pi (\bx -\by)}}~~~~~~~\bG(\vx,\vy) = { 1 \over {\pi (x
-y)}}\cr
}\eqno(16)
$$
where $x,~y$ and $\bx,~\by$ are holomorphic and antiholomorphic variables
respectively.
$p$ acts as the right translation operator on $M$ or $H$ and $\bp$ as the left
translation
operator on $M^\dag$ or $H$, namely
$$\eqalign{
[p_a(\vy),M(\vx)]&=M(\vx) (-it_a) \delta (\vx-\vy)\cr
[\bp_a (\vy), M^\dag (\vx)] &= (-it_a)M^\dag (\vy) \delta (\vx-\vy)\cr}\eqno(17)
$$
$p_a,~\bp _a$ can be written as functional differential operators once a 
parametrization of
$H$ is chosen. 

${\cal{G}}~,~\bar{\cal{G}}$ in (14) are the regularized versions of the Green's
functions
$G(\vx,\vy)$ and $\bG (\vx,\vy)$. They have been defined in ref.~[2] as
$$\eqalign{
\bar{\G} (\vx,\vy)  & =  \bG (\vx,\vy) [ 1 - e^{-|\vx-\vy|^2/\e}
H(x,\by) H^{-1} (y, \by) ] \cr
\G (\vx,\vy)  & =  G (\vx,\vy) [ 1 - e^{-|\vx-\vy|^2/\e} 
H^{-1}(y,\bx) H (y, \by) ] \cr} \eqno(18)
$$
As $\epsilon \rightarrow 0$, for finite $|\vx - \vy|$, (${\cal{G}},~ \bar{\cal{G}}) \rightarrow (G,~\bG$).
Since regularization is needed for calculations involving $T$ or $T'$ we have 
included it
at this stage.

In deriving (14) the following relations expressing the action of the operators
$p_a,~\bp _a$ on the Wess-Zumino actions $\S (M),~\S (M^{\dag})$ were used 
$$\eqalign{
\bp _a \S (M^{\dag}) & = - {i \over 2\pi} \bdel(\del M^{\dag} M^{\dag -1})_a \cr
p _a  \S (M) & = - {i \over 2\pi} \del(M^{-1} \bdel M)_a \cr}
\eqno(19)
$$

Comparing (14) with the corresponding expression for the Hamiltonian of the
pure
Yang-Mills case, we see that we can write ${\cal{H}}'$ as (up to an overall
constant)
$$\eqalign{
{\cal{H}}' &= {\cal{H}}_{YM}  +{ie^2k\over 8\pi} \left[
(\bdel H H^{-1})_a {\bar \G} \bp_a - (H^{-1}\partial H)_a \G p_a\right]\cr
&~~~~~~~~~~~~~~~+{e^2k^2\over 16\pi^2} \Tr (\partial H H^{-1} \bdel H H^{-1})\cr
}
\eqno(20)
$$
Under a parity transformation, $H\rightarrow H^{-1}, p\rightarrow -\bp, 
\bp\rightarrow -p$ and we see that the term in (14) or (20) which involves just
one
factor of $p,\bp$ violate parity conservation, as expected. ${\cal{H}}_{YM}$ is
of
course parity-invariant.

The expression (14) or (20) for $T'$ is suitable for analyzing symmetries and for comparison with
perturbative analysis, but there is an alternative form which is more convenient for our
purposes.
{}From the definition of the electric field operators in (2), we see that
$$
\bE^a \Psi = e^{-{k\over 4\pi}\int A^a\bA^a }
\left( -{i\over 2}{\delta \over \delta A^a}\right) ~\left(  e^{{k\over 4\pi}\int
A^a\bA^a }
\Psi \right) \eqno(21)
$$
Because of the extra factor of $\exp\left({k\over 4\pi}\int A^a\bA^a \right)$
and the
PW identity
$$
\S (H) = \S (M) +\S (M^\dagger ) -{1\over 2\pi} \int A^a\bA^a \eqno(22)
$$
we can simplify various formulae by defining $\chi (H)= \exp(\half k\S (H) )
~\Phi (H)$.
Thus
$$\eqalign{
\Psi &= \exp\left[ {k\over 2}\left( \S (M^\dagger ) -\S (M) +\S (H)
\right)\right]
~\Phi (H)\cr
&= \exp\left[ k\S (M^\dagger ) -{k\over 4\pi} \int A^a\bA^a \right] ~\Phi
(H)\cr}
\eqno(23)
$$
In terms of the $\Phi$ 's, the corresponding Schr\"odinger equation reads
$$
\tilde{\cal{H}} \Phi (H) = E \Phi (H)
\eqno(24)
$$ 
where $\tilde{\cal{H}}= e^{-{k \over 2} \S (H)} {\cal{H}}' e^{{k \over 2}\S (H)}
=
\tilde {T} + V $.
Using (19) we can write $\tilde{T}$ as
$$\eqalign{
\tilde{T} & = {e^2 \over 2} \int H_{ab} (\vx) \left( \int_y \bar{\cal{G}} (\vx,\vy) \bp
_a (\vy)
- {ik
\over 2\pi} (\del H H^{-1})_a \right) \int_u{\cal{G}}(\vx,\vu) p_b (\vu)\cr
& = {e^2 \over 2} \int H_{ab}  e^{- k \S (H)}  \bar{\cal{G}}  \bp _a 
e^{k \S (H)} {\cal{G}} p_b \cr
&= T_{YM} - {ie^2 k \over 4\pi} \int H_{ab} (\del H H^{-1})_a {\cal{G}} p_b \cr}
\eqno(25)
$$

In terms of $\Phi$'s, the inner product is given by
$$
\la 1\vert 2\ra = \int d\mu (H) e^{(k+2c_A)\S (H)}~~\Phi_1^* \Phi_2
\eqno(26)
$$
This inner product for the YMCS theory agrees with what is obtained for the pure
CS theory as
well [8]. Compared to the pure YM case, the essential difference in the measure
is to
replace $2c_A$ by $k+2c_A$ as the coefficient of the WZW action $\S (H)$.

While expression (25) is self-adjoint with respect to the measure (26), it is not
manifestly so.
An
alternative expression for the kinetic energy operator $\tilde{T}$ can be
obtained,
which is manifestly self-adjoint. We can write a general matrix element of
$\tilde{T}$
as
$$
\la 1\vert \tilde{T} \vert 2\ra = {e^2 \over 2}\int d\mu (H) e^{(k+2c_A) \S
(H)} \Phi_1^* ~ H_{ab} ~ e^{- k \S (H)} ~ \bar{\cal{G}}  \bp _a ~
e^{k \S (H)} ~ {\cal{G}} p_b ~ \Phi_2 
\eqno(27)
$$
{}From previous analysis [2] we found that
$$
\left[ \bar{\cal{G}} \bp_a (\vx) ~,~H_{ab} (\vx) e^{2 c_A \S (H)} \right] =0
$$
Using this we can rewrite (27) as
$$
\la 1\vert \tilde{T} \vert 2\ra = {e^2 \over 2}\int d\mu (H) e^{(k+2c_A) \S
(H)} \Phi_1^* e^{-(k+2c_A) \S (H)}(\bar{\cal{G}}\bp)_a H_{ab} e^{(k+2c_A) \S (H)}
({\cal{G}}p)_b \Phi_2
$$
which leads to a self-adjoint expression for $\tilde{T}$ as
$$
\tilde{T}= {e^2\over 2} \int e^{-(k+2c_A) \S (H)}(\bar {\cal{G}} \bp)_a H_{ab} e^{(k+2c_A)
\S (H)}
({\cal{G}} p)_b
\eqno(28)
$$
The expression for $\tilde{T}$ has the same form as in the pure YM theory [2] except for the
$2c_A
\rightarrow k+2c_A$ shift in the coefficient of $\S (H)$. 

In the case of the pure Yang-Mills case we defined the currents
$$
J_a = {c_A \over \pi} \del H H^{-1}~~,~~~~~~~~~\bJ _a = {c_A \over \pi} H^{-1}
\bdel H
\eqno(29)
$$
It turned out that these current operators generated the whole spectrum of the
theory.
The situation is not exactly the same in the case of YMCS theory as we shall
comment
later; however for both YM and YMCS theories, $J_a$ plays the role of the
nonperturbative gluon.

The kinetic energy operator can be simplified in the case of $\Phi$'s being
purely functions of the current $J_a$, rather than $H$ in general; such
a simplification will be
useful later in the evaluation of the vacuum wavefunction. We find
$$\eqalignno{
\tilde{T}&=T_{YM} +{e^2k\over 4\pi} \int J^a {\delta \over \delta J^a}&(30a)\cr
T_{YM}&= {e^2c_A\over 2\pi}\left[ \int_u J^a(\vu) {\d \over \d J^a(\vu)} ~+~
\int
\Omega_{ab} (\vu,\vv) 
{\d \over \d J^a(\vu) }{\d \over \d J^b(\vv) }\right]&(30b)\cr
\Omega_{ab}(\vu,\vv)&= {c_A\over \pi^2} {\d_{ab} \over (u-v)^2} ~-~ 
i {f_{abc} J^c (\vv)\over {\pi (u-v)}}&(30c)\cr}
$$
We see that the coefficient of the $\int J{\delta /\delta J}$-term
is $(k+2c_A)e^2/4\pi$, giving a mass of this value to every factor
of $J^a$, which is consistent with the shift $2c_A\rightarrow 
k+2c_A$.

\vskip .1in
\noindent{\bf 3. Eigenstates of $T$}
\vskip .1in

In analyzing the Schr\"odinger equation and the eigenstates of the Hamiltonian,
we shall follow the strategy we used for the pure Yang-Mills case, namely, we
shall consider eigenstates of $T$ first, neglecting the potential energy
term $V$. This is essentially a strong coupling limit, $e^2 \gg p$, where $p$ is
the typical momentum scale. In the next section we shall see how the effects of
the potential energy term can be included.

The commutation rule (3)
shows that $\bE^a$ is like an annihilation operator while $E^a$ is like a
creation operator. Further, since $T=2e^2\int E^a\bE^a$, we see that the ground
state
or vacuum state for $T$ is given by $T\Psi_0=0$ with $\bE^a\Psi_0=0$. This is 
equivalent to $\Phi=\Phi_0=constant$.
In other words, upto a normalization constant,
$$
\Psi_0= \exp\left[ k\S (M^\dagger ) -{k\over 4\pi} \int A^a\bA^a\right]
\eqno(31)
$$
This is normalizable with the inner product (12).

Since $E^a$ behaves like a creation operator, we should expect that the excited
states can be
obtained by successive applications of $E^a$ on $\Psi_0$. In particular, the
first excited state may be expected to be of the form $E^a\Psi_0$. However, this
is not
gauge invariant. Gauge invariant combinations are given by $E^b(\vx)
M^\dagger_{ab}(\vx)\Psi_0$
and $ M^\dagger_{ab}(\vx) E^b(\vx)\Psi_0$. These can be evaluated as follows.
$$\eqalignno{
M^\dagger_{ab}(\vx) E^b(\vx)\Psi_0 &= {k\over 4c_A}J_a (\vx) \Psi_0&(32a)\cr
E^b(x) M^\dagger_{ab}(\vx) \Psi_0 &= M^\dagger_{ab}(\vx) E^b(\vx) \Psi_0 -{i\over
2}\left[
{\delta M^\dagger_{ab}(\vx)\over \delta \bA^b(\vy)}\right]_{\vy\rightarrow
\vx}\Psi_0&{}\cr
&= \left( {k\over 4c_A}+{1\over 2}\right) J_a (\vx) \Psi_0 = {k+2c_A\over
4c_A}J_a (\vx) \Psi_0&(32b)\cr}
$$
where we have used the fact that, with proper regularization,
$$
 -{i\over 2}\left[
{\delta M^\dagger_{ab}(\vx)\over \delta \bA^b(\vy)}\right]_{\vy\rightarrow \vx}=
\half J_a (\vx)
$$
Thus, apart from constant factors, both $E^bM^\dagger_{ab}$ and $M^\dagger_{ab}
E^b$
involve the current $J_a$. {}From (30), we thus have
$$
T~(M^\dagger_{ab}(\vx) E^b(\vx)\Psi_0)= {e^2\over 4\pi}(k+2c_A) ~(M^\dagger_{ab}(\vx)
E^b(\vx)\Psi_0)
\eqno(33)
$$
It is also useful to work through this directly without first relating it to $J$'s.
We have
$$\eqalign{
T~(M^\dagger_{ab}(\vx) E^b(\vx)\Psi_0)&= 2e^2 \int_y E^k (\vy) [ \bE^k(\vy),
M^\dagger_{ab}(\vx)
E^b(\vx)] \Psi_0\cr
&= {e^2k\over 4\pi} E^b(\vx) M^\dagger_{ab}(\vx) \Psi_0\cr}\eqno(34)
$$
We rewrite this in terms of $M^\dagger E$ using (32). The above equation then 
reduces
to
$$
T~(J_a\Psi_0) = {e^2\over 4\pi}(k+2c_A) (J_a\Psi_0)
\eqno(35)
$$
in agreement with (33). If we start with $E^b(\vx) M^\dagger_{ab}(\vx)\Psi_0$, the
same result
is obtained with a careful treatment of potential singularities.

In the large $k$, or semiclassical limit, the lowest excited state $J_a\Psi_0$ 
has a mass $e^2k/4\pi$
which is the
perturbative mass acquired by the gauge boson due to the CS term.
$J_a\Psi_0$ may thus be identified as the state corresponding to a single gauge
boson
which now has a mass $(k+2c_A)e^2/4\pi$. It is clear that states with many gauge
bosons may be obtained by a suitable product of many $J$'s.

\vskip .1in
\noindent{\bf 4. Potential energy, screening, etc.}
\vskip .1in

We now turn to the inclusion of the potential energy term $V= {1\over 2e^2} \int
B^2$ and how this modifies the vacuum wavefunction.
In the case of the pure YM theory, this could be done in a power series in
$1/e^2$ and the
resulting series summed up to obtain the vacuum wavefunction in the form
$\Phi_0 =e^P$ with
$$
P= -{2\pi^2 \over e^2c_A^2}\int \bdel J_a \left( {1\over
{m+\sqrt{m^2-\nabla^2}}}\right)
\bdel J_a ~+ ~3J-terms \eqno(36)
$$
where $m=e^2c_A/2\pi$. The $3J$-terms involve at least 3 powers of the current
$J_a$. Such terms were shown to be subdominant, compared to the $2J$-term
displayed,
for modes of $J$ in the large and small momentum regimes. For low momentum modes, 
$P$ simplifies and gives
$$
\Phi_0 \approx \exp\left[ -{1\over 2me^2}\int \Tr ~B^2\right]
\eqno(37)
$$
Vacuum expectation values are thus reduced to correlators in a Euclidean
two-dimensional
YM theory of coupling constant $g^2= me^2= e^4c_A/2\pi$. The expectation value
of the Wilson
loop operator then follows an area law. The string tension $\sigma$ was obtained
as
$\sqrt{\sigma}= e^2 \sqrt{(N^2-1)/8\pi}$, which is in very good agreement with
Monte Carlo
estimates [9].

In the present case of YMCS theory, the kinetic energy operator has a structure
similar to the pure YM case, viz.,
$$\eqalign{
\tilde{T} &= {\tilde m}\int_u J^a(\vu) {\d \over \d J^a(\vu)} ~+~ m \int
\Omega_{ab} (\vu,\vv) 
{\d \over \d J^a(\vu) }{\d \over \d J^b(\vv) } \cr
&={\tilde m} \left[ \int_u \xi^a(\vu) {\d \over \d \xi^a(\vu)} ~+~ \int
\Omega_{ab} (\vu,\vv) 
{\d \over \d \xi^a(\vu) }{\d \over \d \xi^b(\vv) }\right]\cr}\eqno(38)
$$
where ${\tilde m}= (k+2c_A)e^2/4\pi$ and $\xi= \sqrt{{\tilde m}/m}~J$. The inclusion of the
potential energy can thus be done in a way similar to the pure 
YM case. For the term with two currents in $\Phi_0$ we get
$\Phi_0=e^P$, where
$P$ is given by (36)
with $m\rightarrow {\tilde m}$ and $J\rightarrow \xi $. In other words, 
$$\eqalign{
\Phi_0&= \exp\left[ -{\pi\over {\tilde m} c_A}\int \bdel \xi\left( {1\over {\tilde m}+
\sqrt{ {\tilde m}^2 -\nabla^2 }}\right) \bdel \xi \right]\cr
&=\exp\left[ -{\pi\over mc_A}\int \bdel J\left( {1\over {\tilde m}+
\sqrt{ {\tilde m}^2 -\nabla^2 }}\right) \bdel J \right]\cr}\eqno(39)
$$ 
For low momentum modes, we
also have the approximation
(37), again with $m\rightarrow {\tilde m}$. We thus find $\vert \Phi_0\vert^2
=e^{-S^{(2)}}$, 
where $S^{(2)}$ corresponds to the action
of a Euclidean two-dimensional YM theory of $g^2 = {\tilde m}e^2 =
e^4(k+2c_A)/4\pi$. 

However, we have more than a 2d-YM theory in the evaluation of correlators.
The volume measure for gauge invariant configurations is $d\mu (H) \exp(2c_A\S
(H))$.
For the YMCS theory, the inner product as given by (26) involves
$d\mu (H) \exp[(k+2c_A)\S (H)]$. Thus the
vacuum expectation value of an operator ${\cal O}$ in the 
$(2+1)$-dimensional YMCS theory can be written, for long wavelength modes, as
$$\eqalign{
\la {\cal O}\ra &= \int d\mu (H) e^{(k+2c_A)\S (H)}~e^{-{1\over 4g^2}\int F^2} ~{\cal O}\cr
&= \int d\mu\left( {\cal{C}}\right)~e^{k\S(H)}e^{-{1\over 4g^2}\int F^2}~
{\cal O}\cr}\eqno(40)
$$ 
The extra factor $e^{k\S}$ can also be expressed in
terms 
of integration
over two-dimensional fermions as
$$\eqalign{
e^{\S (H)}&= \det (D\bD )=\int [dQ]e^{-\int (\bQ_LD Q_L +\bQ_R \bD Q_R )}\cr
&= \int [dQ]~e^{\int \bQ\gamma\cdot D Q}\cr}\eqno(41)
$$
where the $Q$'s are fermions in the fundamental representation of $SU(N)$.
Thus with $k$ flavors of such fermions we get the factor $e^{k\S (H)}$. 
An alternative way of writing (40) is therefore
$$\eqalignno{
\la {\cal O}\ra &= \int [dQ]d\mu (H) e^{2c_A\S (H)}~e^{-{\cal F}}~{\cal
O}&{(42a)}\cr
&= \int [dQ]d\mu\left({\cal{C}}\right)~e^{-{\cal F}}~{\cal O}&{}\cr
{\cal F}&= \int d^2x~\left[ {1\over 4g^2}F^a_{\mu\nu}F^{a\mu\nu} +\sum_{i=1}^k
\bQ^i\gamma
\cdot D Q^i \right]&(42b)\cr}
$$
where $g^2= e^4(k+2c_A)/4\pi$. In other words, the problem is equivalent to the
computation of functional averages in a two-dimensional YM theory coupled to $k$
flavors of massless Dirac fermions in the fundamental representation.

Two-dimensional YM theory coupled to fermions in the fundamental representation
has
been analyzed by a number of authors [10]. Some of their results can be taken over to
the
present context. For example, the presence of massless dynamical fermions leads
to
screening of Wilson loop operators. We may thus conclude that the average of the
Wilson loop operator for the YMCS theory will not obey an area law; rather we
should
have $\la W_C\ra = e^{-w}$, with $w/A_C \rightarrow 0$ as the area $A_C
\rightarrow
\infty$.

The appearance of fermions in the fundamental representation in (42) makes the
screening of charges plausible from an intuitive point of view. Nevertheless,
(42) ultimately provides only a mathematically useful representation; there are,
of course, no fermions in the YMCS theory we are considering. In fact, for the 
expectation value of $W_C$, one can directly
do the integration using (40); the term $e^{k\S}$ leads to short range propagators
for the gauge potentials and hence to screening.

Since
screening effects generally require the presence of charged particles in the physical spectrum,
one might ask the question of whether these can be understood in terms of the gauge fields
themselves. The inner product (26) shows that matrix elements in the YMCS theory are obtained in
terms of a hermitian WZW model of level
$(k+2c_A)$. The correlators in this model are the analytic continuation of the correlators of the
level $k$ $SU(N)$ WZW-model with
$\kappa = k+c_A$ replaced by $-\kappa = -(k+c_A)$. The states of finite norm in
YMCS theory can thus be constructed in terms of the integrable primary operators of
the
$SU(N)$-theory and not just the currents; such primary operators other than the
identity do exist for
$k\neq 0$ [1,2,8]. An example of such a state would be
$$\eqalign{
\alpha &= U(\infty, \vx) M (\vx) \cr
U(\infty, \vx)&= P\exp\left[ \int_{\vx}^\infty A \right] \cr}
\eqno(43)
$$
$\alpha$ is gauge invariant under transformations which go to the identity at
spatial infinity.  Under gauge
transformations which go to a constant $h\neq 1$ at spatial infinity, $\alpha
\rightarrow h\alpha$, which is what is expected of a charged state. 

\vskip .2in
\noindent{\bf 5. Comparison with previous results}
\vskip .1in

As mentioned before, there have been many papers analyzing different aspects of
the 
YMCS theory. A Hamiltonian analysis, which is closest in spirit to ours, is in a
recent
paper by Grignani {\it et al} where a gauge invariant construction of eigenstates of
$T$ is presented [5].
The construction of the states is as follows. Define
$$
\S(g, A,B) =  \S(g) + {1 \over \pi}\int  \Tr  [g^{-1} \bdel g B - \bA \del g
g^{-1} 
+ \bA g B g^{-1}- {1 \over 2} (\bA A + \bB B)]
\eqno(44)
$$
where $g(x)$ is an $SU(N)$-valued field, $\S(g)$ is the WZW action and $B$ is an
auxiliary field
variable. Equation (44) is like a ``gauged" WZW action with $(\bA,~B)$ acting as the
gauge field. Using
(2), we can check immediately that $\bE ^{a} e^{-k \S(g,A,B)} = 0 $.
The strategy used in ref.~[5] is to use this fact and define the vacuum
wavefunction for $T$ as
$$
\Psi _0 [A,B]= \int [dg]~ e^{-k \S (g,A,B)}
\eqno(45)
$$
where the group integration is done with the Haar measure. The resulting
$\Psi_0 [A,B]$ still depends
on the auxiliary variable $B$ but by defining the inner product as
$$
\la 1\vert 2\ra = \int {{[dAdB]} \over {\rm vol {\cal{G_*}}}} ~\Psi_1 ^{*} (A,B) \Psi _2
(A,B)
\eqno(46)
$$
one has a systematic way to compute matrix elements.  (${\cal{G_*}}$ is the space
of all gauge transformations which approach a constant at spatial infinity.) Under a gauge
transformation, $M \rightarrow
h M$, ~ $M^{\dag} \rightarrow M^{\dag}  h^{-1}$, we have $g \rightarrow h g$.
Since $E^{b}(\vx)$
behaves as a ``creation" operator by virtue of (3), one can use the gauge
transformation
property of $g$ and define a gauge invariant excited state
$$
\Psi_a [\vx;A,B] = E_b (\vx) \int [dg]~ g_{ba}(\vx) e^{-k \S (g,A,B)}
\eqno(47)
$$
Higher excited states were also defined in ref.~[5]; for comparison with our
results, the lowest
excited state (47) is adequate to illustrate the main points. Since $\bE$ has
the commutation
rule (3) with $E$ and does not affect $g_{ba}$, it was argued in ref.~[5], that
$\Psi _a [\vx;A,B]$ is an
eigenstate of $T$ with eigenvalue $e^2 k /4 \pi$. It was indicated that the
proper regularization
of $E \bE$ in $T$ might modify the result, although no calculations to this
effect were given. In comparing with our results in section 3 the following
questions arise: How are the states $\Psi _a [\vx;A,B]$ related to our excited
states created by the action of the current $J_a$ on the $\Psi_0$, eq. (32)? If
they are related what is the correct energy eigenvalue? These questions were
already posed in ref.~[5]. In what follows we present an analysis which
essentially provides an answer to these questions and the source of the
discrepancy between our results and the ones in ref.~[5].

We begin with the observation that $\Psi_0 [A,B]$ in (45) is proportional
to our $\Psi_0$ in (31). We can
write $B = -\del N
N^{-1}$ for some $SL(N,{\bf C})$-matrix $N$ and then (44) can be written as 
$$
\S (g,A,B) = \S (M^{\dag} g N) - \S (M^{\dag}) - \S (N) - {1 \over {2 \pi}} 
\int \Tr ( A \bA + B \bB )
\eqno(48) 
$$
Equation (45) now becomes 
$$
\eqalign{
\Psi_0 [A,B] & = e^{k \S (M^{\dag}) -{k \over 4\pi} \int A^a \bA ^a}  e^{k \S (N)-{k
\over 4\pi}\int B^a
\bB ^a}  \int dg e^{-k \S (M^{\dag} g N)} \cr
&= C~e^{k \S (M^{\dag}) - {k \over 4\pi} \int A^a \bA ^a}  \cr}
\eqno(49)
$$
where, by invariance of the Haar measure, $dg=d (M^{\dag}gN)$ and hence $C$ is
independent of $A,
\bA$. $\Psi_0 [A,B]$ in (45) agrees upto a normalization constant with our
$\Psi_0$ in (31).

Consider now the excited state (47). Carrying out the action of $E_b$ on $\S
(g,A,B)$ we find
$$
\Psi _a [\vx; A,B] = {ik \over 4 \pi}  \int [dg]~ g_{ba}(\vx) [A^b-(gBg^{-1} - \del 
g g^{-1})^b](\vx)
e^{-k \S (g,A,B)}
\eqno(50)
$$
There are singularities involved in this expression. $\Psi _a [\vx;A,B]$ involves averages
with the level
$k$ $SU(N)$ WZW model and, in particular, it involves the combination $g_{ba}(\vx)
\del g
g^{-1}(\vx)$. Since $\del g g^{-1}$ is the current of the WZW model, there are
singularities in
the WZW correlator $\la \del g g^{-1}(\vy) g_{ba}(\vx)\ra$ as $\vy \rightarrow \vx$, as is
evident from the
standard operator product expansions. This is also seen from the use of (48) to
simplify (50).
We find
$$
\eqalign{
\Psi_a[\vx;A,B] = {ik \over 4\pi} \int [dg] \big[ &(M^{\dag -1} P N^{-1})_{ba} (\vx)(-\del M
M^{-1} - M^{\dag
-1} \del M^{\dag})_b (\vx)\cr
   & + g_{ba} (\vx) (M^{\dag -1} \del P P^{-1} M^{\dag})^b (\vx) \big]  
   e^{-k \S(g,A,B)} \cr}
\eqno(51)
$$
where $P = M^{\dag} g N$.

The operator product expansion gives
$$
\Tr (t_s \del P P^{-1})(\vy) g_{ba}(\vx) = {-i \over k} {{M^{\dag}_{sl} (\vx) 
f_{lbm}g_{ma}(\vx)}
\over {(y-x)}} +
\cdot
\eqno(52)
$$
We see that (47), as written, is not well defined.
To avoid the singularity in the second term of
(51), we must introduce some sort of
point-splitting. We can define a proper regularized version of (47) by
$$
\Psi_a [\vx;A,B] = R_{cb}(\vx,\vy) E_c(\vx) \int [dg] g_{ba}(\vy) e^{-k \S(g,A,B)} |_{\vy \rightarrow
\vx}
\eqno(53a)
$$
where
$$
R_{cb} (\vx,\vy) = \left[ e^{-A(x-y) - \bA (\bx - \by)} \right] _{cb}
\eqno(53b)
$$
is the Wilson line from $y$ to $x$ (written above for small separation). With this
regularization we can now evaluate $\Psi _a[x;A,B]$ to get
$$
\Psi_a [\vx;A,B]= {{k+2c_A} \over 2c_A} \la (PN^{-1})_{ba}\ra J_b (\vx) \Psi_0
\eqno(54a)
$$
where
$$
\la (PN^{-1})_{ba}\ra  = \int [dP] ~P_{bs} e^{-k \S(P)} N^{-1}_{sa}
\eqno(54b)
$$
We see that $\Psi_a [x;A,B]$ is indeed proportional to our excited state (32). However
two remarks
concerning (54) are in order. First of all, regularization is important, not
just for the
evaluation of $T$, but for defining the $\Psi _a$ itself via (47). Secondly, the
average of $P$ in
(54b) is actually zero. Thus the state would be zero unless this trivial
(independent of $A,~ \bA$)
factor can be removed by some suitable procedure. It may be possible to do this by defining
``renormalized" $g_{ab}$'s with factors depending on infrared and ultraviolet cutoffs
(similar to what is done in minimal conformal models [11]). This problem appears only
in the case of the lowest excited state $\Psi _a [\vx;A,B]$. 

Since $\Psi _a [\vx;A,B] $ is proportional to $J _a (\vx) \Psi _0$ we would expect that 
its energy eigenvalue at the strong coupling limit (its $T$ eigenvalue) should be 
${e^2 \over 4\pi} (k + 2 c_A)$, according to our analysis. Thus regularization
should indeed shift the
eigenvalue from the value $e^2 k /4 \pi$ found in ref.~[5]. We shall now go over the salient
features of such a
regularized calculation. First of all, instead of (53b) we use the more
sytematic expression
$$\eqalign{
R_{cb}(\vx,\vy)&= \left[ M^{\dag -1} (x, \bx ) H(x,\by ) H^{-1}( y,\by )
M^\dag (y,\by )\right]_{cb}
\sigma (\vx,\vy,\epsilon )\cr
\sigma (\vx,\vy,\epsilon )&= {1\over \pi \epsilon}\exp\left({-\vert \vx-\vy\vert^2/ 
\epsilon}\right)
\cr}\eqno(55)
$$
As $\epsilon \rightarrow 0$, $\sigma (\vx,\vy,\epsilon ) \rightarrow \d (\vx, \vy)$ and
$R_{cb}(\vx,\vy) \rightarrow \d_{cb} \d (\vx, \vy)$ as desired.
Expansion of (55) for small separations coincides with (53b). The regularized state
is given by
$$
\Psi_a [\vx;A,B]= \int_y \int [dg] ~R_{cb}(\vx,\vy) E^c(\vx) g_{ba}(\vy) e^{-k\S (g,A,B)}
\eqno(56)
$$
The action of $E$ on $\S (g,A,B)$ produces the following expression for (56)
$$ \eqalign{
& \Psi_a[\vx;A,B] = \cr
& {k \over 2 c_A} \int_y \sigma (\vx,\vy,\epsilon) \int [dg] e^{-k \S(g)}(g N^{-1})_{ba}
(y) [H(x,\by) H^{-1}(y,\by)]_{cb}
      \left[ J - {c_A \over \pi} (\del g g^{-1}) \right]_c (\vx)  \Psi_0 \cr}
\eqno(57)
$$
In applying $T$ on $\Psi_a[x;A,B]$ we encounter the following terms.
$$\eqalign{
T\Psi_a [\vx;A,B] &=I +II +III +IV +V \cr
I&= {ke^2 \over c_A} \int \sigma e^{-k \S} (gN^{-1})_{ba}
   \left[E^k,\left[\bE^k,[HH^{-1}]_{cb} \right]\right]
   \left(J - {c_A \over \pi} (\del g g^{-1}) \right)_c \Psi_0\cr
II&= {ke^2 \over c_A} \int \sigma e^{-k \S} (gN^{-1})_{ba} \left[\bE^k,[HH^{-1}]_{cb} \right]
           [E^k,J_c] ~\Psi_0\cr
III&= {ke^2 \over c_A} \int \sigma e^{-k \S} (gN^{-1})_{ba} \left[\bE^k,[HH^{-1}]_{cb} \right]
   \left(J - {c_A \over \pi} (\del g g^{-1}) \right)_c 
           E^k \Psi_0\cr
IV&= {ke^2 \over c_A} \int \sigma e^{-k \S} (gN^{-1})_{ba} \left[E^k,[HH^{-1}]_{cb} \right]
           \bE^k J_c ~\Psi_0\cr
V&= {ke^2 \over c_A} \int \sigma e^{-k \S} (gN^{-1})_{ba} [HH^{-1}]_{cb}
           E^k \bE^k J_c ~\Psi_0}
\eqno(58)
$$
Term $V$ has been essentially calculated in (35). We find
$$
V = {k \over 2c_A} {e^2(k+2 c_A) \over 4 \pi} J_b \Psi_0
        \int  e^{-k \S (g)} (gN^{-1})_{ba}  \eqno(59)
$$

In evaluating the other terms we make the following observations. The commutator of 
$H(x,\by) H^{-1}(y,\by)$
with $E$ or $\bE$ is of order $(x-y)$ and it vanishes as $x \rightarrow y$. However there are singularities of the order ${1 \over {x-y}}$ 
coming from the
operator product expansion $g_{be}(\vy) (\del g g^{-1})_c(\vx)$ in terms $I$ and $III$. As a
result one might expect a finite contribution from $I$ and $III$. There are also
possible new singularities due to the coincidence limit taken when $\bE(\vw)$
and $E(\vw)$ act on the operators at the same position, which needs more
careful treatment. 
With the regularized expressions for $E,\bE$ from (2), (15) and (18) we have 
$$
[\bE_k(\vw), H(x,\by) H^{-1}(y,\by)] = -{1 \over 2} M_{kl}(\vw)
   [ \G_{lm}(\vw,x,\by) - \G_{lm}(\vw,\vy) ] H(x,\by) T^m H^{-1}(y,\by) 
   \eqno(60)
$$
where $T^m$ are the Lie algebra generators in the adjoint representation.
After tedious but straightforward calculations, the commutator 
$ \int_w [E,[\bE,H H^{-1}]] $ can be shown to be
$$
\int_w [E(\vw), [\bE(\vw), H(x,\by) H^{-1}(y,\by)]] 
    = {1 \over 4} (x-y) J(\vx) + O((x-y)^2) \eqno(61)
$$
Combining this with (52) we get
$$
I = {c_A e^2 \over 2 \pi} J_b \Psi_0 \int  e^{-k \S (g)} (gN^{-1})_{ba} 
\eqno(62)
$$
Similarly, we get a nonzero finite result from term $III$,
$$
III = {k e^2 \over 4 \pi} J_b \Psi_0 \int  e^{-k \S (g)} (gN^{-1})_{ba} 
\eqno(63)
$$
The remaining terms $II$ and $IV$ need careful evaluation; the calculations are
once again quite long since most of the integrals involved do not admit 
approximations for the parameter-regimes of interest. Eventually
the terms $II$ and $IV$ are seen to vanish. Using the results (58-63), we then find
$$\eqalign{
T\Psi_a [\vx;A,B] &= {e^2 (k + 2 c_A) \over 4\pi} {k + 2c_A \over 2c_A} 
               J_b \Psi_0 \int  e^{-k \S (g)} (gN^{-1})_{ba} \cr
        &= {e^2 \over 4 \pi} (k+ 2c_A) \Psi_a[\vx;A,B] }\eqno(64)
$$
The eigenvalue of $T$ is indeed 
$(e^2/4\pi) (k+2c_A)$ in agreement with our results. Thus the analysis of
ref.~[5], if carried further, taking account of regularizations, will give results 
identical to ours.

Similar results hold for the higher excited states considered in [5]. As an
example we consider the properly regularized excited state with two $E$ 's,
namely 
$$\eqalign{
& \Psi_{a_1 a_2}[\vx_1,\vx_2;A,B] = \int [dg] \prod_{i=1}^2 R_{c_i b_i}(\vx_i,\vy_i)
E_{c_i}(\vx_i)  g_{{b_i} a_i}(\vy_i) e^{-k \S(g,A,B)} = \cr
& \left( {k \over {2c_A}} \right)^2 \int \prod_{i=1}^2  
  \sigma(\vx_i,\vy_i,\epsilon) [H(x_i,\by_i)H^{-1}(y_i,\by_i)]_{c_i b_i}
  \left[J - {c_A \over \pi}(\del g g^{-1})
\right]_{c_i} (\vx_i)
  g_{b_i a_i}(\vy_i) \Psi_0 [A,B] \cr}
\eqno(65)
$$
This expression can be simplified by carrying out the operator product expansion as
before. In this case, we need to reduce three- and four-point 
functions of the type
$$ \eqalignno {
& \int e^{-k S(g)} g_{b_1 a_1}(\vy_1) g_{b_2 a_2} (\vy_2)(\del g g^{-1})_{c_1}(\vx_1) 
&(66a) \cr
& \int e^{-k S(g)} g_{b_1 a_1}(\vy_1) g_{b_2 a_2} (\vy_2) 
       (\del g g^{-1})_{c_1}(\vx_1) (\del g g^{-1})_{c_2}(\vx_2) &(66b) \cr}
$$
into two-point functions of two $g$'s. After taking singularities from
operator product expansions into account, we find
$$\eqalign{
\Psi_{a_1 a_2}[\vx_1,\vx_2; A,B] 
  =& \left\{ \left(1+{k \over 2c_A}\right)^2 J_{b_1}(\vx_1) J_{b_2}(\vx_2)
  + \left(1+{k \over 2c_A}\right) \bG(\vx_1,\vx_2)[J(\vx_1) + J(\vx_2)]_{b_1 b_2}
   \right.\cr
   & \left. + {k \over 2} [\bG(\vx_1,\vx_2)]^2 \delta_{b_1 b_2} \right\}
   \int e^{-k \S(g)} (gN^{-1})_{b_1 a_1}(\vy_1) 
                    (gN^{-1})_{b_2 a_2} (\vy_2) \Psi_0}  \eqno(67)
$$
where the factor 1 in the first two terms has come from regulators as in the
lowest excited 
$J$ state. Using the expression (38) of $T$ in terms of $J$, $T$ on 
$\Psi_{a_1 a_2}[\vx_1, \vx_2; A,B]$ becomes
$$
T\Psi_{a_1 a_2} = 2 \tilde{m} \Psi_{a_1 a_2}
                  + 2 \tilde{m} c_A [G(\vx_1,\vx_2)]^2 \delta_{b_1 b_2}
                    \int e^{-k \S(g)} (gN^{-1})_{b_1 a_1}(\vx_1) 
                                     (gN^{-1})_{b_2 a_2}(\vx_2) \Psi_0
\eqno(68)
$$
Therefore the state $\Psi_{a_1 a_2}$ is essentially an eigenstate of $T$ (by
redefining the constant part in $\Psi_{a_1 a_2}$ to absorb the last term
in (68)). The corresponding eigenvalue is $2 {e^2 \over {4 \pi}} (k+2c_A)$. 
In obtaining this result, it is crucial that the ratio of the coefficients of the 
two-$J$ 
and one-$J$ terms in $\Psi_{a_1 a_2}$ is as in (67); otherwise one does not get an
eigenstate of $T$
of the form (38). Although we did not explicitly calculate eigenvalues of $T$ for even
higher excited states, we expect the eigenvalue of the state involving $n~E$'s or
$n~J$'s to be $n {e^2 \over {4 \pi}} (k+2c_A)$. Thus we see that the contribution of the regulator 
used in defining states is important.
An interesting point of the above calculation is that the
eigenvalues add up without any correction, irrespective of the
positions $\vx_1$ and $\vx_2$. Therefore, in the strong coupling limit where the potential
term is neglected, there is no interaction between $J$'s which modifies 
the eigenvalue. In other words, the $J$'s can be far separated from each other. 
This is consistent with the fact that there is no
confinement in YMCS theory. One might ask the question of whether such states
can also occur in pure YM theory. We expect these states to be
nonnormalizable in this case, in agreement with the expected confinement in the pure
Yang-Mills case.

Our calculation which gives the shift $2c_A \rightarrow k+2c_A$ takes account of the
nonperturbative corrections to the mass. This is
different from the shift $k\rightarrow k+c_A$ expected from perturbative calculations
[12]. Such perturbative corrections to our calculation may also exist; 
they can in principle be calculated using perturbation theory in the Hamiltonian formalism.
The explicitly self-adjoint expression (28) for $\tilde{T}$ is most suited for this analysis;
the mixing of parity-violating and conserving terms, as in the calculation of
Pisarski and Rao, should produce the perturbative corrections.
Cornwall has recently pointed out that the inclusion of both perturbative and nonperturbative
effects for the mass can perhaps lead to some sort of critical behaviour at a certain value
of $k$ [13]. Our calculations do not go far enough to make more precise statements
regarding this question.
\vskip .2in
\noindent {\bf 6. Conclusion}
\vskip .1in
We have performed a Hamiltonian analysis of Yang-Mills theory with a level $k$
Chern-Simons term in terms of gauge invariant variables. The low lying spectrum of
this theory at the strong coupling limit has been obtained.
The gauge invariant version of the massive
gauge boson states is
given in terms of the current $J^a$. The mass of the gauge boson is ${e^2 \over
4\pi}(k+2 c_A)$.
Long distance properties of vacuum expectation values are related to
a Euclidean
two-dimensional YM theory coupled to $k$ flavors of Dirac fermions in the
fundamental
representation. The expectation value of the Wilson loop operator should exhibit
screening rather
than an area law. Related comments and
comparison with previous results are also given.
\vskip .2in
\noindent{\bf Acknowledgements}
\vskip .1in

This work was supported in part by the NSF grant PHY-9605216.
CK thanks Lehman College of CUNY and Rockefeller University for hospitality
facilitating the completion of this work.

\vskip .2in
\noindent{\bf References}
\vskip .1in
\item{1.}
D. Karabali and V.P. Nair, {\it Nucl.Phys.} {\bf B464} (1996) 135; {\it Phys.Lett.}
{\bf B379} (1996) 141; {\it Int. J. Mod. Phys} {\bf A12} (1997) 1161.
\item{2.}
D. Karabali, Chanju Kim and V.P. Nair, {\it Nucl.Phys.} {\bf B524} (1998) 661.
\item{3.} D. Karabali, Chanju Kim and V.P. Nair, {\it Phys.Lett.} {\bf B434} (1998)
103.
\item{4.} R. Jackiw and S. Templeton, {\it Phys.Rev.} {\bf D23} (1981) 2291;
J. Schonfeld, {\it Nucl.Phys.} {\bf B185} (1981) 157; S. Deser, R. Jackiw
and
S. Templeton, {\it Phys.Rev.Lett.} {\bf 48} (1982) 975; {\it Ann.Phys.}
{\bf 140} (1982) 372.
\item{5.} G. Grignani, G. Semenoff, P. Sodano and O. Tirkkonen, {\it Nucl.
Phys.} {\bf B489} (1997) 360.
\item{6.}  E. Witten, {\it Commun.Math.Phys.} {\bf 92} (1984) 455;
S.P. Novikov, {\it Usp.Mat.Nauk} {\bf 37} (1982) 3.
\item{7.} A.M. Polyakov and P.B. Wiegmann, {\it Phys.Lett.} {\bf B141} (1984) 223.
\item{8.} K. Gawedzki and A. Kupiainen, {\it Phys.Lett.} {\bf B215} (1988) 119;
{\it Nucl.Phys.} {\bf B320} (1989) 649;
M. Bos and V.P. Nair, {\it Int.J.Mod..Phys.} {\bf A5} (1990) 959.
\item{9.} M. Teper, {\it Phys. Lett.} {\bf B311} (1993) 223; hep-lat/9804008 and
references therein.
\item{10.} D.J. Gross, I.R. Klebanov, A.V. Matytsin and A.V. Smilga, {\it Nucl. Phys.}
{\bf B461} (1996) 109; Y. Frishman and J. Sonnenschein, {\it Nucl. Phys.} {\bf B496}
(1997) 285
; D. Kutasov and A. Schwimmer, {\it Nucl. Phys.} {\bf B442} (1995) 447.
\item{11.} V.S. Dotsenko and V.A. Fateev, {\it Nucl. Phys.} {\bf B240} (1984) 312.
\item{12.} R.D. Pisarski and S.. Rao, {\it Phys.Rev.} {\bf D32} (1985) 2081.
\item{13.} J.M. Cornwall, {\it Phys. Rev.} {\bf D54} (1996) 1814; {\it Int. J. Mod.
Phys.} {\bf A12} (1997) 1023.

\bye